\documentclass[12pt,amsbsy]{article}
\newcommand{\beq}{\begin{equation}}
\newcommand{\eeq}{\end{equation}}
\newcommand{\beqn}{\begin{eqnarray}}
\newcommand{\eeqn}{\end{eqnarray}}

\newcommand{\om}{\mbox{${\omega}$}}

\begin{document}



\begin{center}
{\bf \large Quasinormal Modes, Quantized Black Holes,\\

and Correspondence Principle}
\end{center}

\begin{center}
I.B. Khriplovich\footnote{khriplovich@inp.nsk.su}
\end{center}
\begin{center}
Budker Institute of Nuclear Physics\\
630090 Novosibirsk, Russia,\\
and Novosibirsk University
\end{center}

\bigskip

\begin{abstract}
Contrary to the wide-spread belief, the correspondence principle
does not dictate any relation between the asymptotics of
quasinormal modes and the spectrum of quantized black holes.
Moreover, this belief is in conflict with simple physical
arguments.
\end{abstract}

\bigskip


The quasinormal modes (QNM) are the solutions of the homogeneous
wave equation in the gravitational field of a black hole with the
boundary conditions corresponding to outgoing waves at spatial
infinity and incoming waves at the horizon. Two boundary
conditions make the frequency spectrum $\om_n$ of QNMs, so-called
ringing frequencies, discrete. Of course, the solutions for QNMs
are not finite everywhere. We discuss here the gravitational
perturbations of the Schwarzschild black hole.

The calculation of $\,\rm{Im}\,\om_n$ for $n \gg 1$ in the
semiclassical approximation reduces to a relatively simple
exercise in complex plane, with the result
\beq\label{iom}
M\,\rm{Im}\, \om_n = \, -\, \frac{1}{4}\,
\left(n+\,\frac{1}{2}\right)
\eeq
(here $M$ is the black hole mass; the gravitational constant $k$
and the velocity of light $c$ are put to unity). However the
corresponding problem for $\,\rm{Re}\,\om_n$ turned out much more
difficult. This asymptotics was found initially in Refs. [1--3] by
numerical methods:
\beq\label{rom0}
M\,\rm{Re}\, \om_n = 0.0437123, \quad n\gg 1.
\eeq
Then a curious observation was made in Ref. \cite{hod}: the result
(\ref{rom0}) can be presented as
\beq\label{rom}
M\,\rm{Re}\, \om_n = \,\frac{\ln 3}{8\pi}\,,
\eeq
or
\beq\label{ht}
\rm{Re}\, \om_n = \,\frac{\ln 3}{8\pi M}\,= \, T_H\,\ln 3,
\eeq
where $T_H$ is the Hawking temperature. Expression (\ref{rom}) for
the asymptotics of $\rm{Re}\, \om_n$ was afterwards derived in
Refs.\cite{mot,mon} analytically.

Certainly, the investigations of ringing frequencies for
Schwarzschild black holes, as well as for Reissner-Nordstr\o m and
Kerr ones, are of a real interest by themselves.

Meanwhile, it was suggested in Ref. \cite{hod} that the asymptotic
value (\ref{rom}) for $\,\rm{Re}\,\om_n$ of the gravitational
perturbation is of crucial importance for the quantization of
Schwarzschild black holes. The idea has become quite popular and
attempts are made to apply it to the quantization not only of
Schwarzschild black holes, but of charged and rotating ones as
well. There are scores of papers on the subject, and their list is
too lengthy to be presented in this short note.

These attempts, starting with Ref. \cite{hod}, appeal essentially
to the correspondence principle quoted in Ref. \cite{hod} (with
reference to \cite{bor}) as follows: ``transition frequencies at
large quantum numbers should equal classical oscillation
frequencies''.

It goes without saying that this principle is correct. However,
its exact meaning is as follows \cite{ll}. In a quantized system,
the distance $\Delta E$ between two neighbouring levels with large
quantum numbers differing by unity, i. e. between levels with $n$
and $n+1$ ($n \gg 1$), is related to the classical frequency $\om$
of this system by relation
\beq
\Delta E = \hbar \om.
\eeq
In other words, in the semiclassical case $n \gg 1$, the
frequencies corresponding to transitions between energy levels
with $\Delta n \ll n$ are integral multiples of the classical
frequency $\om$.

So, contrary to the assumption of Ref. \cite{hod}, in the
discussed problem of a black hole, large quantum numbers $n$ of
the correspondence principle are unrelated to the asymptotics
(\ref{ht}) of ringing frequencies, but are quantum numbers of the
black hole itself. Thus, the belief that the area spacing of a
black hole is expressed via asymptotics (\ref{ht}), despite its
great popularity, in no way follows from the respectable
correspondence principle.

Moreover, this belief is in conflict with simple arguments. The
reasoning is as follows. The real part of ringing frequencies does
not differ appreciably from its asymptotic value (\ref{rom0}) in
the whole numerically investigated range of $n$, starting from $n
\sim 1$. Meanwhile the imaginary part grows as $n+1/2$, and
together with it the spectral width of quasinormal modes (in terms
of common frequencies) also increases linearly with $n$. In this
situation, the idea that the resolution of a quasinormal mode
becomes better and better with the growth of $n$, and that in the
limit $n \to \infty$ this mode resolves an elementary edge (or
site) of a quantized surface, does not look reasonable.

\begin{center}***\end{center}
I am grateful to V.F. Dmitriev and V.M. Khatsymovsky for
discussions, to L. Smolin for comments in correspondence, and to
the anonymous referee for urging to make my criticism more
concrete. The investigation was supported by the Russian
Foundation for Basic Research through Grant No. 03-02-17612.

\end{document}